\documentclass[iop]{emulateapj}
\usepackage{epsfig}
\usepackage{apjfonts}
\usepackage{color}
\usepackage{aas_macros}

\usepackage{amsfonts}
\usepackage{amssymb}
\usepackage{amsmath}
\usepackage{graphicx}
\usepackage{epsfig}
\usepackage{color}

\def\apj{ApJ}
\def\apjl{ApJ}
\def\apjs{ApJS}
\def\ao{Appl.~Opt.}

\def\aap{A\&A}

\def\mnras{MNRAS}

\def\prd{Phys.~Rev.~D}

\def\physrep{Phys.~Rep.}

\def\nar{New~Astro.~Rev.}

\def\sun{\odot}

\shorttitle{Signature of a New-Born Black Hole}
\shortauthors{Chen et al.}

\begin{document}

\title{Signature of a New-Born Black Hole from the Collapse of a Supra-massive Millisecond Magnetar}

\author{Wei Chen$^{1}$, Wei Xie$^{2}$, Wei-Hua Lei*$^{1}$, Yuan-Chuan Zou$^{1}$, Hou-Jun L\"{u}$^{3}$, En-Wei Liang$^{3}$, He Gao$^{4}$ and Ding-Xiong Wang$^{1}$}
\affil{$^{1}$School of Physics, Huazhong University of Science and Technology, Wuhan 430074, China. Email: leiwh@hust.edu.cn
}
\affil{$^{2}$School of Physics and Electronic Science, Guizhou Normal University,  Guiyang, 550001, China.}
\affil{$^{3}$Guangxi Key Laboratory for Relativistic Astrophysics, Department of Physics, Guangxi University, Nanning 530004, China.}
\affil{$^{4}$Department of Astronomy, Beijing Normal University, Beijing 100875, China.}

\begin{abstract}
The X-ray plateau followed by a steep decay (``internal plateau'') has been observed in both long and short gamma-ray burst (GRBs), implying a millisecond magnetar operating in some GRBs. The sharp decay at the end of plateau, marking the abrupt cessation of the magnetar central engine, has been considered as the collapse of a supra-massive magnetar to a black hole (BH) when it spins down. If ``internal plateau'' is indeed the evidence of a magnetar central engine, a natural expectation is a signature from the new-born BH in some candidates. In this work, we find that GRB 070110 is a particular case, which shows a small X-ray bump following its ``internal plateau''. We interpret the plateau with a spin-down supra-massive magnetar and the X-ray bump with a fall-back BH accretion. This indicates that the new-born BH is likely active in some GRBs. Therefore, GRB 070110-like events may provide a further support to the magnetar central engine model and enable us to investigate the properties of the magnetar as well as the new-born BH.
\end{abstract}

\keywords{accretion, accretion disks - black hole physics - gamma-ray burst: individual (GRB 070110) }

\section{Introduction}
Long gamma-ray bursts (GRBs) are likely related to the core-collapse of massive stars \citep{W93, P98, MW99}. Short GRBs have been proposed to originate from the merger of two neutron stars (NS-NS, \citep{E89, NPP92}) or the merger of a neutron star and a black hole (NS-BH, \citep{P91}). However, the nature of the central engine of GRBs remains unknown.

Recently, modeling various afterglow features for both long and short GRBs within the framework of the millisecond magnetar (or pulsar with weaker magnetic field) central engine model has gained growing attention \citep{Dai06, M08, D11, Fan11, B12, B13, G14, LZ14, Lv15}. In particular, the so-called ``internal X-ray plateaus'' with rapid decay at the end of the plateaus, are difficult to be interpreted within the framework of a BH central engine, but are consistent with a rapidly spinning millisecond magnetar as the central engine. The abrupt decay is naturally understood as the collapse of a supra-massive magnetar into a BH after the magnetar spins down \citep{T07, R10, R13, Z14}.

Actually, ``internal plateaus'' have been discovered in both long and short GRBs \citep{T07, LZZ07, Ly10, R10, R13, LZ14, Lv15}. If magnetars are indeed the central engine of GRBs, and the sudden drop after the plateau are interpreted as the collapse of a supra-massive NS into a BH, then the signatures from this new-born BH are expected. Especially for long GRBs, the giant X-ray bumps, likely due to fall-back accretion onto a BH, have been discovered, e.g., in GRB 121027A and GRB 111209A \citep{Wu13, Yu15, Gao16a}. These observations imply that a fraction of the envelope materials would fall back and activate the accretion onto the BH at late times \citep{kumar08a, kumar08b}. Therefore, a direct expectation is that there should be GRBs with ``internal plateau'' followed by an X-ray bump.

In this paper, we find that GRB 070110 is a potential candidate of this kind. The small X-ray bump following the ``internal plateau'', first uncovered by \cite{T07}, is likely the result of a fall-back accretion. Therefore, GRB 070110 may exhibit a clue to the new-born BH from the collapse of a supra-massive magnetar, and provides us a good opportunity to study the properties of this new-born BH. In Section~\ref{sec:TheModel} , we give a general picture of the spin-down of a supra-massive magnetar and the fall-back accretion onto the new-born BH. We then apply the model to GRB 070110 in Section~\ref{sec:GRB070110}. In Section~\ref{sec:Conclusion}, we briefly summarize our results and discuss the implications. Throughout the paper, a concordance cosmology with parameters $ H_0=71 ~\rm km \ s^{-1} Mpc^{-1}$, $\rm \Omega_M = 0.30$, and $\Omega_\Lambda =0.70$ is adopted.

\section{The Model}
\label{sec:TheModel}
Magnetar has been proposed to be the candidate central engine of some GRBs. There are several scenarios involved in this engine to explain the prompt emission: 1) the hyper-accretion onto the NS as presented in \cite{ZD09, ZD10} and \cite{B13}; 2) magnetic bubbles launched in a differentially millisecond proto-NS \citep{Dai06}; 3) as suggested by \cite{Li16}, the post-merger product of NS-NS mergers might be fast-rotating supramassive quark stars (QSs) rather than NSs, therefore, the prompt emission may be powered by the phase transition of a QS \citep{CD96}; 4) protomagnetar wind model as proposed by \cite{M11}.

\begin{figure}
\begin{center}
\begin{tabular}{ll}
\resizebox{80mm}{!}{\includegraphics[]{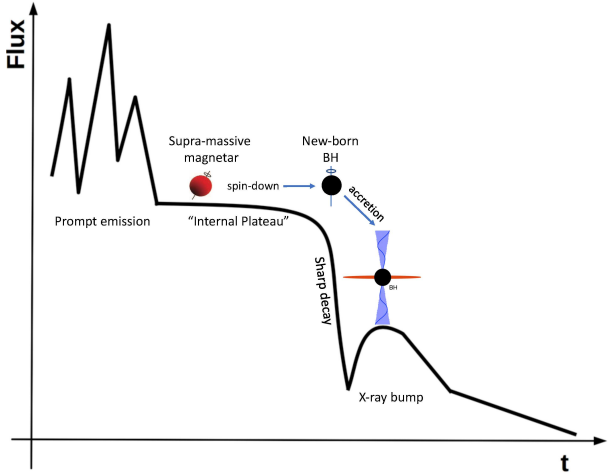}}
\end{tabular}
\caption{
Illustration of the expected ``internal plateau'' observations with signature of the new-born BH. The ``internal plateau'' is powered by the spin-down power of a supra-massive magnetar. The steep decay marks the collpase of the magnetar into a BH when it spins down. The fall-back accretion onto the BH may produce a X-ray bump at the end of steep decay, making it a signature of the new-born BH.}
\label{fig:model}
\end{center}
\end{figure}

A supra-massive magnetar may collapse into a BH after it spins down. The observational correspondence of this prediction is the ``internal X-ray plateau'' discovered in some long and short GRBs \citep{T07, Ly10, R10, R13, LZ14, Lv15}. The successful interpretations of the giant X-ray bumps in GRBs, like GRB 121027A and GRB 111209A, imply that a fraction of the envelope materials would fall back and activate the accretion onto the BH at late time \citep{kumar08a, kumar08b, Wu13, Yu15, Gao16a}. One thus expects the observation of an X-ray bump/flare following the ``internal X-ray plateau'' as the signature the new-born BH. The expected lightcurve and the sketch of the model are illustrated in Figure~\ref{fig:model}.

\subsection{``Internal plateau'' powered by a spin-down supra-massive magnetar}
\label{sec:InternalPlateau}
``Internal plateaus'' are suggested to be powered by a spin-down supra-massive Magnetar. The characteristic spin-down luminosity $L_0$ and the characteristic spin-down time scale $\tau$ are related to the magnetar initial parameters as \citep{ZM01}
\begin{equation}
 L_0 = 1.0 \times 10^{49}~{\rm erg~s^{-1}} (B_{p,15}^2 P_{0,-3}^{-4} R_6^6),
\label{L0}
\end{equation}
\begin{equation}
 \tau = 2.05 \times 10^3~{\rm s}~ (I_{45} B_{p,15}^{-2} P_{0,-3}^2 R_6^{-6}),
\label{tau}
\end{equation}
where $I_{45}$ is the moment of inertia in units of $10^{45} ~\rm g\ cm^2$, $B_{p,15}$ is the magnetic field strength in units of $10^{15}~ \rm G$, $P_{0,-3}$ is the initial period in milliseconds, and $R_6$ is stellar radius in units of $10^6$ cm. The convention $Q = 10^x Q_x$ is adopted in cgs units for all other parameters throughout the paper.

The spin-down luminosity $L_0$ is related to the plateau luminosity ($L_{\rm b, iso}$) as,
\begin{equation}
\eta_{\rm X} L_0 = L_{\rm b, iso} f_{\rm b},
\label{Lb}
\end{equation}
where $\eta_{\rm X}$ is the radiation efficiency, and $f_{\rm b}=1-\cos\theta_{\rm j}$ is the beaming factor.

The spin-down formula due to dipole radiation is given by \citep{S83},
\begin{equation}
P(t) = P_0 (1+ \frac{t}{\tau})^{1/2}.
\end{equation}

A supra-massive magnetar is temporarily supported by rigid rotation, and it collapses to a BH at a later time when it spins down. The collapse occurs when the maximum gravitational mass $M_{\rm max}$ becomes equal to the total gravitational mass of the protomagnetar $M_{\rm NS}$. Here, the critical mass $M_{\rm max}$ depends on the magnetar spin period $P$ as \citep{L03},
\begin{equation}
M_{\rm max} = M_{\rm TOV} (1+ \hat{\alpha} P^{\hat{\beta}}),
\end{equation}
where $M_{\rm TOV}$ is the maximum mass for a nonrotating NS, $\hat{\alpha}$ and $\hat{\beta}$ rely on NS Equation of State (EoS). \cite{L14} worked out the numerical values for $M_{\rm TOV}$, the NS radius ($R$), the moment of inertia ($I$), and thus $\hat{\alpha}$ and $\hat{\beta}$ for several EoSs. Recent studies with short GRB data favor the EoS GM1 ($M_{\rm TOV}=2.37 ~M_\sun$, $R=12.05 ~\rm km$, $I=3.33 \times 10^{45}~ \rm g\ cm^{-2}$, $\hat{\alpha}=1.58\times 10^{-10} {\rm s}^{-\hat{\beta}}$ and $\hat{\beta} = -2.84$), if we assume that the cosmological NS-NS merger systems have the same mass distribution as the observed Galactic NS-NS population \citep{Lv15, Gao16b}. Our modeling is not sensitive to the choice of EoS. For simplicity,  we just adopt EoS GM1 in the paper.

The collapse time of magnetar into BH, $t_{\rm col}$, can be generally identified as the observed break time, i.e. $t_{\rm col} = t_{\rm b}/(1+z)$. Since the post-plateau decay slope is much smaller than $-2$, the spin-down timescale should be greater than the break time. We thus take $t_{\rm b}/(1+z)$ as the lower limit of the spin down timescale.

\subsection{The X-ray bump due to fall-back accretion onto the new-born BH}
The new-born BH is expected to be active. Especially in long GRBs, the progenitor star has a core-envelope structure, as is common in stellar models. The core part collapses into a rapidly spinning supra-massive magnetar, and the envelope mass falls back to the new-born BH after the collapse of NS.

The evolution of the fall-back accretion rate are described with a broken-power-law function of time as \citep{M01, Zhang08, Dai12, Chevalier89}

\begin{eqnarray}
\dot{M} = \dot{M}_{\rm p} \left[ \frac{1}{2}\left(\frac{t-t_0}{t_{\rm p}-t_0} \right)^{-1/2} +  \frac{1}{2}\left(\frac{t-t_0}{t_{\rm p}-t_0} \right)^{5/3} \right]^{-1},
\label{dotm}
\end{eqnarray}
where $t_0$ is the beginning time of fall-back accretion in the local frame. We also assume a turn-off time of the BH central engine $t_{\rm f}$, beyond which the tail emission from jet fades as $t^{-(2+\beta)}$, where $\beta$ is the spectral index.

The hyper-accreting BH system can launch a relativistic jet via neutrino-antineutrino annihilation \citep{P99, NPK01, D02, Gu06, CB07, J04, J07, Liu07, Liu15, Lei09, Lei17, XLW16}, or Blandford-Znajek mechanism (hereafter BZ; \citep{BZ77, Lee00, Li00, Lei05, Lei13}). The neutrino annihilation mechanism suffers strong baryon loading from the disk and therefore may be too ``dirty'' to account for a GRB jet \citep{Lei13, XLW17}. For this reason, we suppose that the jet powering the X-ray bump is dominated by the BZ mechanism.

For a Kerr BH with mass $M_\bullet (\equiv m_{\bullet} M_\odot)$ and angular momentum $J_\bullet$, the BZ power is \citep{Lee00,Li00,Wang02,Mckinney05,Lei08,Lei11,Lei13}
\begin{equation}
L_{\rm BZ}=1.7 \times 10^{50} a_{\bullet}^2 m_{\bullet}^2
B_{\bullet,15}^2 F(a_{\bullet}) \ {\rm erg \ s^{-1}},
\label{eq_Lmag}
\end{equation}
where $a_\bullet=J_\bullet c/(GM_\bullet^2)$ is the spin parameter of the BH, $B_{\bullet}$ is the magnetic field strength threading the BH horizon, and $F(a_{\bullet})=[(1+q^2)/q^2][(q+1/q) \arctan q-1]$, of which $q= a_{\bullet} /(1+\sqrt{1-a^2_{\bullet}})$.

Considering that the BH magnetic field is supported by the surrounding disk, one can estimate its value by equating the magnetic pressure on the horizon to the ram pressure of the accretion flow at its inner edge \citep{Moderski97},
\begin{equation}
\frac{B_{\bullet}^2}{8\pi} = P_{\rm ram} \sim \rho c^2 \sim \frac{\dot{M} c}{4\pi r_{\bullet}^2},
\label{Bmdot}
\end{equation}
where $r_{\bullet}=(1+\sqrt{1-a_\bullet^2})r_{\rm g}$ is the radius of the BH horizon, and $r_{\rm g} = G M_\bullet /c^2$. Then the BZ power can be rewritten as a function of mass accretion rate as
\begin{equation}
L_{\rm BZ}=9.3 \times 10^{53} \frac{a_\bullet^2 \dot{m}  F(a_\bullet)}{(1+\sqrt{1-a_\bullet^2})^2} \ {\rm erg \ s^{-1}} ,
\label{eq:EB}
\end{equation}
where $\dot{m} \equiv \dot{M}/(M_\sun s^{-1})$ is the dimensionless accretion rate.

The observed X-ray luminosity is connected to the BZ power via the X-ray radiation efficiency $\eta$ and the jet beaming factor $f_{\rm b}$, i.e.,
\begin{equation}
\eta_{\rm X} L_{\rm BZ}=f_{\rm b} L_{\rm X,iso},
\label{eqX}
\end{equation}
here, $\eta_{\rm X}$ and $f_{\rm b}$ use the same values as in Section~\ref{sec:InternalPlateau}.

BH would be spun-up by accretion while spun-down by the BZ mechanism. The evolution equations are given by \citep{Wang02},
\begin{equation}
\frac{dM_\bullet c^2}{dt} = \dot{M} c^2 E_{\rm ms} - L_{\rm BZ},
\label{eq:dM}
\end{equation}

\begin{eqnarray}
\frac{da_\bullet}{dt} = \frac{(\dot{M} l_{\rm ms} - T_{\rm BZ})c}{GM_\bullet ^2} - \frac{2 a_\bullet (\dot{M} c^2 E_{\rm ms} - L_{\rm BZ})}{M_\bullet c^2},
\label{eq:da}
\end{eqnarray}
where $T_{\rm BZ}$ is the BZ torque with expression,
\begin{eqnarray}
T_{\rm BZ} =3.36 \times 10^{45} a_\bullet^2 q^{-1} m_{\bullet}^3 B_{\bullet,15}^2  F(a_\bullet){\rm \ g \ cm^2 \ s^{-2}}. \nonumber \\
\end{eqnarray}

In Equations (\ref{eq:dM}) and (\ref{eq:da}), $E_{\rm ms}$ and $l_{\rm ms}$ are the specific energy and angular momentum at the radius of the innermost stable circular orbit $r_{\rm ms}$ of the disk, respectively, which are defined as \citep{NT73}
\begin{equation}
l_{\rm ms} = \frac{G M_\bullet}{c} \frac{2 (3 \sqrt{R_{\rm ms}} -2 a_\bullet) }{\sqrt{3} \sqrt{R_{\rm ms}} },
\end{equation}

\begin{equation}
E_{\rm ms} = \frac{4\sqrt{ R_{\rm ms} }-3a_{\bullet}}{\sqrt{3} R_{\rm ms}},
\end{equation}
where $R_{\rm ms} = r_{\rm ms}/r_{\rm g} =  3+Z_2 -\left[(3-Z_1)(3+Z_1+2Z_2)\right]^{1/2}$, for $0\leq a_{\bullet} \leq 1$ \citep{B72}, where $Z_1 \equiv 1+(1-a_{\bullet}^2)^{1/3} [(1+a_{\bullet})^{1/3}+(1-a_{\bullet})^{1/3}]$, and $Z_2\equiv (3a_{\bullet}^2+Z_1^2)^{1/2}$.

\section{GRB 070110}
\label{sec:GRB070110}
GRB 070110, with duration $T_{90}  (15-150 ~{\rm keV}) = 89\pm 7$ s and redshift $z=2.352\pm$ 0.001, was triggered by the \textit{Swift} BAT on 2007 January 10 \citep{T07}. The fluence over the 15-150 keV band was $1.8^{+0.2}_{-0.3} \times 10^{-6} ~\rm erg \ cm^{-2}$, from which one can derive an observed isotropic energy of $E_{\gamma, \rm iso} \simeq 3.1 \times 10^{52} ~\rm erg$ \citep{T07, Du16}.

\begin{figure}
\begin{center}
\begin{tabular}{ll}
\resizebox{80mm}{!}{\includegraphics[]{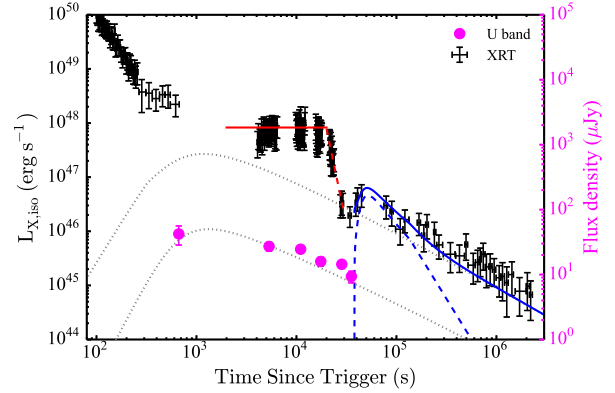}}
\end{tabular}
\caption{Modeling results for the XRT and optical lightcurve of GRB 070110.
The observed data are exhibited with black (XRT, using the data from the Swift data archive) and magenta (optical U-band, the data were taken from \cite{T07}) points with error bar, and the theoretical modeling are shown with red (``internal plateau'' phase) and blue (X-ray bump phase) solid lines. The thin black and magenta dotted lines denote the external shock component in X-ray and optical band, respectively. The blue dashed line corresponds to the contribution from BZ jet according to the fall-back BH-disk.}
\label{fig:fit}
\end{center}
\end{figure}

The afterglow of GRB 070110 \citep{T07} showed a near flat plateau ($\alpha \sim 0.05$) with a flux $L_{\rm b, iso} \sim 10^{48} ~\rm erg \ s^{-1}$ in 0.3-10 keV. It extends to over $10^4$ s before rapidly falling off with a decay index $\alpha \sim 9$ (see the red dashed line in Figure~\ref{fig:fit}). Such a rapid decay cannot be accommodated in any external shock model, so that the entire X-ray plateau emission has to be attributed to internal dissipation of a central engine wind \citep{LZ14, Du16}.

After the sharp decay following the plateau, a small bump was observed \citep{T07}. In \cite{T07}, the fast rise of GRB 070110, after the abrupt drop at $t_{\rm b} \simeq 1.4 \times 10^4$ s, and the following decay has been well described by a fast-rise-exponential-decay (FRED) profile, peaking at $t_{\rm p}(1+z) \simeq  5 \times 10^4$ s. The late X-ray ($>t_{\rm f} =10^5$ s) exhibited a power-law decay with $\alpha \sim 0.7$, which can be interpreted as the standard external forward shock afterglow. To exhibit the significance of the bump, we do empirical function fit to the late-time ($t \geq 1.4 \times 10^4$ s) X-ray light curve with two distinct models, i.e., a single power-law (SPL) function and an FRED bump+PL function. Our results show that the fit with FRED+PL model is significantly better than the one with a SPL model (the reduced chi-square (p-values) are 2.7 (0.00021) for the FRED+PL model, and 10.03 (0.18152) for the SPL model). Considering that FRED+PL model will introduce extra parameters comparing with the PL model, we adopt Bayesian information criterion(BIC) to evaluate the goodness of the two models \citep{S78}. BIC can be written as: $BIC = n \log(RSS/n) + k \log(n)$, where $k$ is the number of free parameters, $n$ is the number of data points, and $RSS$ is the residual sum of squares. As suggested by \cite{S78}, the model with lowest BIC is preferred. We find that the BIC values are -77.48 for the FRED+PL model and -72.71 for the SPL model. Therefore, from statistical point of view, modelling with a FRED like bump is more consistent with the data.

The optical afterglow was detected by Swift/UVOT in the White, U, B, and V filters. The UVOT lightcurves show a decaying behavior that can be well described by a simple power law. It was found that the UVOT late slopes ($\alpha_{opt} \sim 0.63$) are consistent, within the errors, with the late-time X-ray slope. The late optical/X-ray spectrum ($t = 100$ ks) can be described by a continuous power law (with spectral index $\beta = 1.00 \pm 0.14$), indicating that the optical and the late X-ray afterglow may arise from the same physical component.

In Figure~\ref{fig:fit}, the X-ray and optical lightcurve of GRB 070110 are presented with black and magenta points, respectively. An ``internal X-ray plateau'' and a small X-ray bump are identified, making GRB 070110 a possible candidate to study the magnetar and the new-born BH.

Now we apply our model to GRB 070110.

The X-ray plateau could be interpreted with the magnetar model by using Equations (\ref{L0}) - (\ref{Lb}). Since no jet break feature was observed, \cite{Du16} estimated a lower limit of the jet opening angle $\theta_{\rm j}$ with the last observed point ($t_{\rm j} \sim $25 days) in X-ray afterglow, i.e., $\theta_{\rm j} > 7.4^{\rm o}$. In our calculation, $\theta_{\rm j} \simeq 10^{\rm o}$ (and thus $f_{\rm b} \simeq 0.02$) is taken. In fact, the modeling results for the afterglow component depend weakly on the values of $\theta_{\rm j}$.

We adopt a numerical code developed for external shock model \citep{Gao13, Wang14} to model both the late-time ($>10^5$s) X-ray afterglow and the U-band optical lightcurves, as shown with gray (X-ray) and magenta (optical) dotted lines in Figure~\ref{fig:fit}. The empirical function fitting to the optical lightcurve by \cite{T07} suggests a peak at ~1000s, which is usually taken as the jet deceleration time $t_{\rm dec}$. Since $t_{\rm dec} \propto E_{\rm k, iso}^{1/3} n^{-1/3} \Gamma_0^{-8/3}$, the values of jet isotropic kinetic energy $E_{\rm k,iso}$, ambient medium density $n$ and initial bulk Lorentz factor $\Gamma_0$ are chosen to satisfy $t_{\rm dec} \sim 1000 \rm s$. Besides, the optical to X-ray spectral index $\beta = 1.00 \pm 0.14$ suggests an electron energy spectral index of $p\sim 2$. Other microphysics shock parameters, such as $\epsilon_{\rm e}$ and $\epsilon_{\rm B}$, are taken the values so the model can fit both the X-ray and optical flux. However, these model parameters still suffer degeneracy when fitting the X-ray and optical data \citep{kumar15}. For the purpose of this work, we do not attempt to fit the data across a large parameter space. In Table~\ref{tab:parametes}, we present a set of afterglow parameter values that could interpret the data well.

The radiation efficiency $\eta_{\rm X}$ is unknown due to the lack of knowledge on jet dissipation process during plateau and X-ray bump phases. In our calculations, we take $\eta_{\rm X} \sim 0.1$ as a typical value.

Assuming the spin-down timescale $\tau = t_{\rm b}/(1+z)$ and adopting EoS GM1, one can infer a magnetar initial period of $P_0 \sim 8.2$ ms and a magnetic field $B_{\rm p} \sim 4.95 \times 10^{15} $ G from the data, as shown in Table~\ref{tab:parametes}. The mass of the supra-massive magnetar is $M_{\rm NS} \sim 2.37 ~M_{\sun}$, which equals the critical mass $M_{\rm max}$ at collapse time $t_{\rm col}$. The predicted spin-down luminosity of magnetar is also shown with red solid line in Figure~\ref{fig:fit}. The red dashed line in Figure~\ref{fig:fit} is drawn directly with the the empirical function $L_{\rm X, iso} = L_{\rm b, iso} (t/t_{\rm b})^{-9}$.

The FRED-shape X-ray bump in GRB 070110, implying the restart of central engine after the sudden drop of ``internal plateau'', is not predicted in the previous supra-massive magnetar model.

Bumps in the GRB afterglow are often interpreted as due to density variations in the circumburst medium \citep{DL02, DaiWu03}. However, this model predicts a very smooth lightcurve as shown in \cite{DaiWu03} and \cite{U14}, which is inconsistent with the fast-rise-shape bump in GRB 070110. After the collapse of supra-massive magnetar, the magnetic flux will be ejected based on the no-hair theorem of BH, which leads to a short duration flare activity (in radio band, it might be a fast radio burst) just at the end of the plateau  \citep{Z14}. These features are quite different than those in GRB 070100. Furthermore, the total magnetic energy in this model is small compared with the GRB energy, so it would have no significant imprint in the X-ray lightcurve. In this paper, we attribute it to the fall-back accretion onto the new-born BH.

The initial set-up for BH can be obtained by assuming that the new-born BH inherits the mass and angular momentum from the supra-massive magnetar. With the above parameters for magnetar, we therefore get the initial BH mass $M_\bullet^{\rm i} \sim 2.37 ~M_\sun $, and initial spin $a_\bullet^{\rm i} \sim 0.04$ (by using $J_\bullet= 2\pi I/P_0$). Then we calculate the time evolution of the BZ power, and compare it with the observations of X-ray bump. We use the same radiation efficiency $\eta_{\rm X}$ and beaming factor $f_{\rm b}$ as those for plateau. The calculation starts at $t_0 = 4.8\times 10^4/(1+z)$ s.

The modeling of the X-ray bump is exhibited in Figure~\ref{fig:fit} (with blue dashed line). The blue solid line denotes the total emission by including the contributions from both BZ jet (blue dashed line) and the external shock (gray dotted line). The modeling parameters are summarized in Table~\ref{tab:parametes}.

The peak accretion rate is $\dot{M}_{\rm p} \sim 1.0 \times 10^{-5} ~M_\sun$. The total accreted mass should be $M_{\rm acc} \sim 0.085 ~M_\sun$. During the accretion, the BH mass increases from $2.37 ~M_\sun$ to $2.45 ~M_\sun$, and the spin from 0.04 to 0.15.

The X-ray bump appears at $\sim 48000$ s after the GRB trigger, which, divided by $1+z$, corresponds to $t_{\rm fb} \sim t_0 \sim 1.4 \times 10^4$ s. This suggests that the minimum radius around which matter starts to fall back is $r_{\rm fb} \simeq 3.7 \times 10^{11} (M_\bullet /2.5~ M_\sun)^{1/3} ({t_{\rm fb}/14000\;\rm s})^{2/3} \rm cm$, which is consistent with the typical radius of a Wolf-Rayet star.

\begin{table*}
\begin{center}{
\caption{Values of parameters adopted for interpreting the broadband data of GRB\,070110.}
\label{tab:parametes}
\begin{tabular}{ccccccc} \hline\hline
 \multicolumn{7}{c}{Mangetar and BH parameters}\\
  \hline
   $M_{\rm NS}~({M_\sun})$ & $B_{\rm p, 15}$     & $P_{0, -3}$    &$M^i_\bullet~(M_{\odot})$ &  $a^i_\bullet$   & $M_{\rm acc}~({ M_{\odot}})$ &$\dot{M}_p~({ M_{\odot}} \rm s^{-1})$ \\
   $2.37$ &$4.95$     &  $8.2$         & $2.37$     &  $0.04$         &$0.085$  &$1.0\times10^{-5}$ \\
   \hline
  \multicolumn{7}{c}{GRB afterglow parameters}\\
  \hline
   $E_{\rm k, iso}~({\rm erg})$     &$\Gamma_0$               &  $n~(\rm{cm^{-3}})$& $\theta~({\rm rad})$&$\epsilon_e$ &  $\epsilon_B$&  $p$  \\
   $1.58\times 10^{54}$     &  $95$ &  $0.13$   & $10$& $0.02$     &  $0.0002$&  $2.01$        \\
  \hline
  \multicolumn{7}{c}{Other parameters}\\
  \hline
   & $t_{\rm b}~({\rm s})$ & $t_0~({\rm s})$ & $t_{\rm p}~({\rm s})$ & $t_{\rm f}~({\rm s})$& $\eta_{\rm X}$ & $f_{\rm b}$\\
   & $140000$ & $48000/(1+z)$ &   $50000/(1+z)$ &   $10^{5}/(1+z)$  & 0.1 & 0.02 \\
   \hline\hline
 \end{tabular}
 }
\end{center}
\end{table*}

\section{Conclusion and discussion}
\label{sec:Conclusion}
``Internal plateaus'' in GRB afterglows are commonly interpreted with a supra-massive magnetar \citep{T07, Ly10, LZ14, Lv15, Gao16b}. The sudden steep decay implies the collapse of a supra-massive magnetar into a BH. To check this model, we expect a signature of the new-born BH from the observation.

GRB 070110, a long burst ($T_{90} \sim 90$ s), is one typical example with ``internal plateau'' \citep{T07}. Interestingly, a small X-ray bump emerges at the end of the rapid decay of internal plateau, suggesting the reactivation of central engine after the collapse of supra-massive NS. Such a bump is beyond the prediction of magnetar central engine model, but can be well interpreted with the fall-back accretion onto a BH. Our work implies that some GRBs are still active after the collapse of magnetar, and showing possible signatures of the new-born BH.

\begin{figure}
\begin{center}
\begin{tabular}{ll}
\resizebox{80mm}{!}{\includegraphics[]{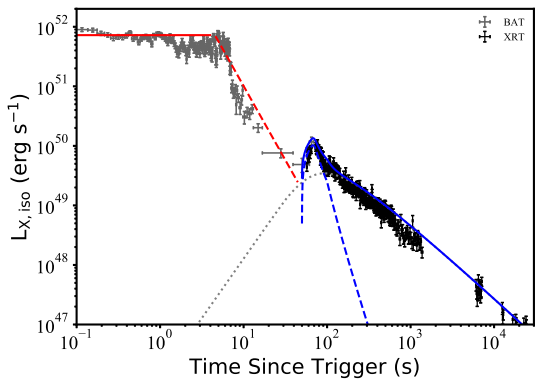}}
\end{tabular}
\caption{Modeling results for the XRT lightcurve of GRB 110731A.
The observed data are exhibited with gray (\textit{swift}/BAT, 15keV-150keV) and black (XRT, 0.3keV-10keV) points with error bar (the data were taken from \cite{Lv17}). The theoretical modeling are shown with red (``internal plateau'' phase) and blue (X-ray bump phase) solid lines. The thin black dotted line denotes the external shock component in X-ray band. The blue dashed line corresponds to the contribution from BZ jet according to the fall-back BH-disk.}
\label{fig:fit2}
\end{center}
\end{figure}

GRB 070110 is not the only burst showing such activity of new-born BH. As shown in Figure \ref{fig:fit2}, GRB 110731A, a ``rest-frame short'' GRB detected by Fermi and Swift observatories at redshift of $z = 2.83$ \citep{T11}, may serve as another example. From the combination of BAT and XRT data, an internal plateau is identified with break time $t_{\rm b} \sim 6$ s. At the end of the steep decay, a fast rise appeared at $\sim 30$ s. Considering the short period of plateau, GRB 110731A may represent a different evolution path of the magnetar. The progenitor of GRB 110731A is probably a compact-star merger \citep{Lv17}. The new-born supra-massive NS would be initially differentially rotating. Within a time scale of seconds, the combination of magnetic breaking and viscosity would drive the star to the uniform rotation phase. The interpretation with our model indicates that the magnetar would have a period of $P_0 \sim 1.3$ ms and a magnetic field of $B_{\rm p} \sim 5.0 \times 10^{16}$. If the initial spin period in this rigid-rotation phase is larger than the critical period supporting the supra-massive NS by centrifugal force, the magnetar would promptly collapse to a BH with initial mass $M_\bullet^{\rm i} \sim 2.37 ~M_\sun $ and initial spin $a_\bullet^{\rm i} \sim 0.2$. Some of the surrounding matter initially blocked by the magnetic barrier of magnetar would begin to fall back and be accreted by the new-born BH, which in turn produced the observed X-ray bump at $t\sim 30$s. The accreted mass should be $\sim 0.002M_\sun$ based on our model. Considering that the typical mass of the ejecta during a merger is in the range $10^{-4} - 10^{-2} M_\sun$ \citep{H13}, we would expect a few percent of them to be accreted by the new-born BH.

If magnetars are the central engines of some GRBs, then several evolution results of magnetars are expected: (1) immediate collapse into a BH; (2) a supra-massive magnetar collapses to a BH at a later time after the it spins down; and (3) a stable magnetar \citep{LZ14, Lv15, Gao16b}. Our analysis suggests that some magnetars may indeed collapse into a BH, which provides a direct support the magnetar central engine model.

However, most GRBs with ``internal plateau'' do not show clear BH signature. The fall-back processes would be intrinsically weak, especially at late time. The bounding shock responsible for the associated supernova for long GRBs, and the outflow from the central engine, would transfer kinetic energy to the surrounding materials. If the injected energy from supernova or central engine is larger, which might be the majority of cases, the fall-back process is vanished.

For simplicity, we didn't include rotational energy loss of the supra-massive magnetar due to gravitational wave (GW) emission. As discussed in \citep{Fan13, Lv15, Gao16b}, a significant energy may be released in the form of GWs. We should investigate this in detail in future.

In this paper, we adopt neutron star EoS GM1 in the modeling. Recently, \citep{Li16} suggested that the ``internal plateau'' might be produced by a quark star. Our model is not sensitive to the EoS, therefore our main results will still hold even if we adopt the QS EoSs.

\section*{Acknowledgements}

We thank the anonymous referee for his/her valuable comments and constructive suggestions. We thank Zi-Gao Dai, Xue-Feng Wu and Bing Zhang for helpful discussions. We acknowledge the use of the public data from the Swift data archive, and the UK Swift Science Data Center. The Numerical calculations were performed by using a high performance computing cluster (Hyperion) of HUST. This work is supported by the National Basic Research Program ('973' Program) of China (grants 2014CB845800), the National Natural Science Foundation of China (U1431124, 11773010, 11603006, 11533003 and U1731239). L.H.J. and E.W. L also acknowledges support by the Guangxi Science Foundation (2016GXNSFCB380005) and special funding for Guangxi distinguished professors (Bagui Yingcai \& Bagui Xuezhe). HG acknowledges support by the National Natural Science Foundation of China under grants 11722324, 11603003, 11690024 and 11633001.


\begin{thebibliography}{24}

\bibitem[Bardeen et al. (1972)]{B72} Bardeen, J.~M., Press, W.~H., \& Teukolsky, S.~A. 1972, ApJ, 178, 347
\bibitem[{Blandford \& Znajek}(1977)]{BZ77} Blandford, R.~D., \& Znajek, R.~L., 1977, MNRAS, 179, 433
\bibitem[Bucciantini et al. (2012)]{B12} Bucciantini, N., Metzger, B.~D., Thompson, T.~A., \& Quataert, E. 2012, MNRAS, 419, 1537
\bibitem[Bernardini et al. (2013)]{B13} Bernardini, M.~G., Campana, S., Ghisellini, G., et al. 2013, ApJ, 775, 67
\bibitem[{Chen \& Beloborodov}(2007)]{CB07} Chen, W.~X., \& Beloborodov, A.~M. 2007, ApJ, 657, 383
\bibitem[Cheng \& Dai (1996)]{CD96} Cheng, K. S., \& Dai, Z. G. 1996, Phys. Rev. Lett. 77, 1210
\bibitem[{Chevalier}(1989)]{Chevalier89} Chevalier, R.~A.\ 1989, \apj, 346, 847
\bibitem[Dai \& Lu (2002)]{DL02} Dai, Z.~G., \& Lu, T. 2002, ApJ, 565, L87
\bibitem[Dai \& Wu (2003)]{DaiWu03} Dai, Z.~G., \& Wu X.-F.\ 2003, \apj, 591, L21
\bibitem[{Dai et al.}(2006)]{Dai06} Dai, Z.~G.,Wang X.~Y.,Wu X.~F., \& Zhang B. 2006, Sci, 311, 1127
\bibitem[{Dai \& Liu}(2012)]{Dai12} Dai, Z.~G., \& Liu R.-Y.\ 2012, \apj, 759, 58
\bibitem[{Dall'Osso et al.}(2011)]{D11} Dall'Osso, S., Stratta, G., Guetta, D., et al. 2011, A\&A, 526, A121
\bibitem[{Di Matteo et al.}(2002)]{D02} Di~Matteo, T., Perna, R. \& Narayan, R. 2002, ApJ, 579, 706
\bibitem[Du et al.(2016)]{Du16} Du, S., L\"{u}, H.~J., Zhong, S.~Q., Liang, E.~W. 2016, MNRAS, 462, 2990
\bibitem[Eichler et al.(1989)]{E89} Eichler, D., Livio, M., Piran, T., \& Schramm, D.~N. 1989, Nature, 340, 126
\bibitem[Fan et al.(2011)]{Fan11} Fan, Y.~Z., Zhang, B.~B., Xu, D., Liang, E.~W., \& Zhang, B. 2011, ApJ, 726, 32
\bibitem[Fan et al.(2013)]{Fan13} Fan, Y.~Z., Yu, Y.~W., Xu, D., et al. 2013, ApJL, 779, L25
\bibitem[Gao et al.(2013)]{Gao13} Gao, H., Lei, W.~H., Zou, Y.~C., Wu, X.~F., \& Zhang, B.\ 2013, \nar, 57, 141
\bibitem[Gao et al.(2016a)]{Gao16a} Gao, H., Lei, W.~H., You, Z.~Q., \& Xie, W. 2016a, ApJ, 826, 141
\bibitem[Gao et al.(2016b)]{Gao16b} Gao, H., Zhang, B., \& L\"{u}, H.~J. 2016b, Phys. Rev. D 93, 044065
\bibitem[Gompertz et al. (2014)]{G14} Gompertz, B.~P., O'Brien, P.~T., \& Wynn, G.~A. 2014, MNRAS, 438, 240
\bibitem[{Gu et al.}(2006)]{Gu06} Gu, W.~M., Liu, T., \& Lu, J.~F. 2006, ApJL, 643, L87
\bibitem[{Hotokezaka et al.}(2013)]{H13} Hotokezaka, K., Kiuchi, K., Kyutoku, K., et al. 2013, PhRvD, 87, 024001
\bibitem[{Janiuk et al.}(2004)]{J04} Janiuk, A., Perna, R., Di~Matteo, T. \& Czerny, B. 2004, MNRAS, 355, 950
\bibitem[{Janiuk et al.}(2007)]{J07} Janiuk, A., Yuan, Y., Perna, R. \& Di Matteo, T. 2007, ApJ, 664, 1011
\bibitem[Kumar et al.(2008a)]{kumar08a} Kumar, P., Narayan, R., \& Johnson, J.~L.\ 2008a, \mnras, 388, 1729
\bibitem[Kumar et al.(2008b)]{kumar08b} Kumar, P., Narayan, R., \& Johnson, J.~L.\ 2008b, Science, 321, 376
\bibitem[Kumar \& Zhang.(2015)]{kumar15} Kumar, P., \& Zhang, B. \ 2015, \physrep, 561, 1
\bibitem[Lasky et al. (2014)]{L14} Lasky, P.~D., Haskell, B., Ravi, V., Howell, E.~J., Coward, D.~M., 2014, PhRvD, 89, 047302
\bibitem[Lee et al.(2000)]{Lee00} Lee, H.~K., Wijers, R.~A.~M.~J., \& Brown, G.~E.\ 2000, \physrep, 325, 83
\bibitem[Lei et al.(2005)]{Lei05} Lei, W.~H., Wang, D.~X., \& Ma, R.~Y. 2005, ApJ, 619, 420
\bibitem[Lei et al.(2008)]{Lei08} Lei, W.~H., Wang, D.~X., Zou, Y.~C. \& Zhang, L. 2008, ChJAA, 8, 404
\bibitem[Lei et al.(2009)]{Lei09} Lei, W.~H., Wang, D.~X., Zhang, L., et al. 2009, \apj, 700, 1970
\bibitem[Lei \& Zhang(2011)]{Lei11} Lei, W.~H., \& Zhang, B.\ 2011, \apjl, 740, L27
\bibitem[Lei et al.(2013)]{Lei13} Lei, W.~H., Zhang, B., \& Liang, E.~W.\ 2013, \apj, 765, 125
\bibitem[{Lei et al.}(2017)]{Lei17} Lei, W.~H., Zhang, B., Wu, X.~F., \& Liang, E.~W. 2017, arXiv: 1708.05043
\bibitem[Li(2000)]{Li00} Li, L.~X.\ 2000, \prd, 61, 084016
\bibitem[Li et al.(2016)]{Li16} Li, A., Zhang, B., Zhang, N.~B., Gao, H., Qi, B., Liu, T. 2016, Phys. Rev. D, 94, 083010
\bibitem[Liang et al. (2007)]{LZZ07} Liang, E.~W., Zhang, B.~B., \& Zhang, B. 2007, ApJ, 670, 565
\bibitem[{Liu et al.}(2007)]{Liu07} Liu, T., Gu, W.~M., Xue, L., \& Lu, J.~F. 2007, ApJ, 661, 1025
\bibitem[Liu et al.(2015)]{Liu15} Liu, T., Hou, S.~J., Xue, L., Gu, W.~M. 2015, ApJS, 218, 12
\bibitem[{L\"{u} \& Zhang}(2014)]{LZ14} L\"{u}, H.~J., \& Zhang, B. 2014, ApJ, 785, 74
\bibitem[{L\"{u} et al.}(2015)]{Lv15} L\"{u}, H.~J., Zhang, B., Lei, W.~H., Li, Y., Lasky, P.~D., 2015, ApJ, 805, 89
\bibitem[{L\"{u} et al.}(2017)]{Lv17} L\"{u}, H.~J., Wang, X.~G., Lu, R.~J., et al. 2017, ApJ, 843, 114
\bibitem[Lyford et al. (2003)]{L03} Lyford, N.~D., Baumgarte, T.~W., \& Shapiro, S.~L. 2003, ApJ, 583, 410
\bibitem[Lyons et al. (2010)]{Ly10} Lyons, N., O'Brien, P. T., Zhang, B., et. al., 2010, MNRAS, 402, 705
\bibitem[MacFadyen \& Woosley (1999)]{MW99} MacFadyen, A.~I., \& Woosley, S.~E. 1999, ApJ, 524, 262
\bibitem[MacFadyen et al.(2001)]{M01} MacFadyen, A.~I., Woosley, S.~E., \& Heger, A.\ 2001, \apj, 550, 410
\bibitem[McKinney(2005)]{Mckinney05} McKinney, J.~C.\ 2005, \apjl, 630, L5
\bibitem[Metzger et al. (2008)]{M08} Metzger, B.~D., Quataert, E., \& Thompson, T.~A. 2008, MNRAS, 385, 1455
\bibitem[Metzger et al. (2011)]{M11} Metzger, B.~D., Giannios, D., \& Thompson, T.~A, et al. 2011, MNRAS, 413, 2031
\bibitem[Moderski et al.(1997)]{Moderski97} Moderski, R., Sikora, M., \& Lasota, J.~P.\ 1997, Relativistic Jets in AGNs, 110
\bibitem[Narayan et al. (1992)]{NPP92} Narayan, R., Paczy\'{n}ski, B., \& Piran, T. 1992, ApJ, 395, L83
\bibitem[{Narayan et al.}(2001)]{NPK01} Narayan, R., Piran, T., \& Kumar, P. 2001, ApJ, 557, 949
\bibitem[Novikov \& Thorne (1973)]{NT73} Novikov, I.~D., \& Thorne, K.~S. 1973, in Black Holes, ed. C. DeWitt-Morette \& B.~S. DeWitt (New York: Gordon \& Breach), 345
\bibitem[Paczy\'{n}ski (1991)]{P91} Paczy\'{n}ski, B. 1991, Acta Astron., 41, 157
\bibitem[Paczy\'{n}ski (1998)]{P98} Paczy\'{n}ski, B. 1998, ApJL, 494, L45
\bibitem[Popham et al. (1999)]{P99} Popham, R., Woosley, S.~E., \& Fryer, C. 1999, ApJ, 518, 356
\bibitem[Rowlinson et al. (2010)]{R10} Rowlinson, A., O'Brien, P.~T., Tanvir, N.~R., et al. 2010, MNRAS, 409, 531
\bibitem[Rowlinson et al. (2013)]{R13} Rowlinson, A., O'Brien, P.~T., Metzger, B.~D., Tanvir, N.~R., \& Levan, A.~J. 2013, MNRAS, 430, 1061
\bibitem[Schwarz et al. (1978)]{S78} Schwarz, G. et al. 1978, The Annals of Statistics, 6, 461
\bibitem[Shapiro \& Teukolsky (1983)]{S83} Shapiro, S., \& Teukolsky, S. 1983, Black Holes, White Dwarfs, and Neutron Stars: the Physics of Compact Objects (New York: Wiley)
\bibitem[Troja et al. (2007)]{T07} Troja, E., Cusumano, G., O'Brien, P.~T., et al., 2007, ApJ, 665, 599
\bibitem[Tanvir et al. (2011)]{T11} Tanvir, N.~R., Wiersema, K., Levan, A.~J., et al., 2011, GCN, 12225
\bibitem[Uhm \& Zhang (2014)]{U14} Uhm, Z. L., \& Zhang, B. 2014, ApJ, 789, 39
\bibitem[Wang et al.(2002)]{Wang02} Wang, D.~X., Xiao, K., \& Lei, W.~H.\ 2002, \mnras, 335, 655
\bibitem[{Wang et al.}(2014)]{Wang14} Wang, J.~Z., Lei, W.~H., Wang, D.~X., et al. 2014, ApJ, 788, 32
\bibitem[Wu et al.(2013)]{Wu13} Wu, X.~F., Hou, S.~J., \& Lei, W.~H.\ 2013, \apjl, 767, L36
\bibitem[Woosley (1993)]{W93} Woosley, S.~E. 1993, ApJ, 405, 273
\bibitem[{Xie et al.}(2016)]{XLW16} Xie, W., Lei, W.~H., \& Wang, D.~X. 2016, ApJ, 833, 129
\bibitem[{Xie et al.}(2017)]{XLW17} Xie, W., Lei, W.~H., \& Wang, D.~X. 2017, ApJ, 838, 143
\bibitem[Yu et al. (2015)]{Yu15} Yu, Y.~B., Wu, X.~F., Huang, Y.~F., Coward, D. M., Stratta, G., Gendre, B., Howell, E.J. 2015, MNRAS, 446, 3642
\bibitem[Zhang (2013)]{Z13} Zhang, B. 2013, ApJL, 763, L22
\bibitem[Zhang (2014)]{Z14} Zhang, B. 2014, ApJL, 780, L21
\bibitem[Zhang \& M\'{e}sz\'{a}ros (2001)]{ZM01} Zhang, B., \& M\'{e}sz\'{a}ros, P. 2001, ApJL, 552, L35
\bibitem[{Zhang et al.}(2008)]{Zhang08} Zhang, W., Woosley, S.~E., \& Heger, A.\ 2008, \apj, 679, 639
\bibitem[Zhang \& Dai (2009)]{ZD09} Zhang, D., \& Dai, Z. G. 2009, ApJ, 703, 461
\bibitem[Zhang \& Dai (2010)]{ZD10} Zhang, D., \& Dai, Z. G. 2010, ApJ, 718, 841

\label{lastpage}

\end{thebibliography}
\end{document}